\documentstyle[12pt,aasms4]{article} 
\def\etal{et al. }
\begin{document}
\def\lapp{\ifmmode\stackrel{<}{_{\sim}}\else$\stackrel{<}{_{\sim}}$\fi}
\def\gapp{\ifmmode\stackrel{>}{_{\sim}}\else$\stackrel{>}{_{\sim}}$\fi}

\title{Parallax of PSR J1744--1134 and the Local Interstellar Medium}

\author
{ M. Toscano$^{1,2}$, M. C. Britton$^2$, R. N. Manchester$^{3,1}$, 
M. Bailes$^2$, J. S. Sandhu$^4$, S. R. Kulkarni$^4$, S. B. Anderson$^4$}
\bigskip
\bigskip
\noindent
\begin{footnotesize}
\noindent
$^1$Physics Department, University of Melbourne, 
Parkville, Vic 3052, Australia; mtoscano@physics.unimelb.edu.au.\\ 
$^2$Astrophysics and Supercomputing, Mail No. 31, Swinburne University 
of Technology, PO Box 218, Hawthorn, Vic 3122, Australia; 
mbritton@pulsar.physics.swin.edu.au, mbailes@swin.edu.au.\\
$^3$Australia Telescope National Facility, CSIRO, PO Box
76, Epping, NSW 2121, Australia; rmanches@atnf.csiro.au. \\
$^4$Department of Astronomy, Caltech, Pasadena CA 91125; jss@astro.caltech.edu, 
srk@astro.caltech.edu, sba@srl.caltech.edu.
\end{footnotesize}
\bigskip
\bigskip

\abstract{ We present the annual trigonometric parallax of 
PSR J1744$-$1134 derived from an analysis of pulse times of arrival. 
The measured parallax, $\pi=$ 2.8$\pm$0.3 mas ranks among the most precisely 
determined distances to any pulsar. The parallax distance of 357$^{+43}_{-35}\,$pc 
is over twice that derived from the dispersion measure using the Taylor \& Cordes 
model for the Galactic electron distribution.  The mean  electron density in the 
path to the pulsar, $n_{e}=$ (8.8$\pm$0.9)$\times$10 $^{-3}$ cm$^{-3}$, is the lowest 
for any disk pulsar. We have  compared the $n_{e}$ for PSR J1744$-$1134 with 
those for another 11 nearby pulsars with independent distance estimates.  We 
conclude that there is a striking asymmetry in the distribution of electrons in 
the local interstellar medium. The electron column densities for pulsars in the 
third Galactic quadrant are found to be systematically  higher than for 
those in the first. The former correlate with the  position of the
well known local H{\small I} cavity in quadrant three.  The excess
electrons within the cavity may be in the form of H{\small II} clouds
marking a region of interaction between the local hot bubble and a 
nearby superbubble. }


\newpage 

\section{Introduction}  

Among the fundamental building blocks of astronomy is the ability to 
measure reliable distances to objects of interest.  These not only provide 
us with their location but also allow us to study properties that are functions 
of distance. Independent and consistent measurements of distances to local 
objects establish a distance  scale upon which all greater distances are based. 
Precise, model-independent measurements of distances to pulsars are rare, yet
they are an invaluable resource with which to study the interstellar
medium (ISM). These pulsar distances provide information about
the column density of free electrons along different lines-of-sight (LOSs).  

Since their discovery, pulsars have had their distances estimated
primarily by measuring the dispersion delay between pulses arriving at
two widely spaced frequencies. As this delay is a function of the integral 
of electron density along the LOS to the pulsar, or dispersion measure (DM), 
a model for the Galactic free electron distribution yields the pulsar's 
distance (e.g., \cite{tc93}). Pulsar distances have also been inferred from 
associations with  objects  with independently measured distances, e.g. globular 
clusters (\cite{lyn95}), supernova remnants (\cite{kas96}), and the Magellanic clouds 
(\cite{fea91}). When detection  of neutral  hydrogen (H{\small I}) absorption of 
the pulsar signal is possible, an estimate, or at least a limit on the distance 
may be obtained using a Galactic rotation model (e.g., \cite{jkww96}).

The most direct estimate of distance to a pulsar is obtained from 
measurement of its annual trigonometric parallax, $\pi$. In
practice, however, measuring $\pi$ is difficult since it requires
astrometric measurements with sub-milliarcsecond accuracy made over
several years. To date parallaxes have been measured for only the 12 
nearby pulsars listed in Table 1. Many of these parallaxes have been 
obtained using long baseline interferometers (e.g., \cite{gwi84}).
In this Letter, we describe a measurement of the parallax of PSR 
J1744$-$1134 obtained by analysing pulse times of arrival (TOAs)
spanning 4 years and combine this result with other similar measurements 
to study the electron distribution in the local ISM (LISM).

\section{Observations, Analysis and Results}

The timing measurements of PSR J1744$-$1134 were made as part of an 
ongoing millisecond pulsar (MSP) timing project using the 64-m Parkes 
radio telescope. We describe briefly the observational and analytic 
methodology used, while a detailed account is presented by Toscano \etal 
(1999)\nocite{tsb+99}.

Between 1995 January and 1999 January we made regular observations of
PSR J1744$-$1134 at 0.66 and 1.4 GHz. At 0.66 GHz we used a dual
linear polarization receiver. During the period of 1997 April until 
1998 August we used the center beam of the Parkes Multibeam receiver 
system for dual linear polarization observations about 1.4 GHz. At 
other times observations at 1.4 GHz were made with a dual circular 
polarization H-OH receiver. The downconverted signal was fed 
into the Caltech correlator (\cite{nav94}) where it was digitized and 
autocorrelated. The autocorrelation functions were folded at the 
topocentric pulse period, Fourier transformed, and compressed to 180
s sub-integrations, 8 frequency channels and 512 phase bins. A
typical observation consisted of 8 contiguous sub-integrations. At 
0.66 GHz, the signal was recorded over 32 MHz of bandwidth. Near 1.4 
GHz we observed with two 128 MHz bands centered near 1.4 and 1.6 GHz 
respectively.

Profiles for each observation were formed by summing all frequency
channels and sub-integrations. The highest signal-to-noise ratio (SNR)
profiles were added to form standard profiles for each observing
frequency. After eliminating profiles with very low SNR and/or high
levels of radio frequency interference we were left with 158 24-min
integrated profiles. Pulse TOAs and TOA errors were determined by
fitting these integrated profiles with the standard profile (\cite{tay90}). 
Arrival-time data were fitted to a pulse timing model using the {\small TEMPO} 
program (http://pulsar.princeton.edu/tempo/). The JPL DE200 ephemeris 
(\cite{sta82}) was used to transform TOAs to the solar system barycenter.

The amplitude and functional form of the residuals from a least-squares 
fit to the TOAs of pulsar position, proper motion, pulse period and its 
first time derivative, and DM, with $\pi$ set to zero is referred to as 
the `timing signal' for parallax. The signal has a semi-annual period and 
amplitude $r^{2} \cos^{2} \theta / (2cd)$, where $r$ is the Earth's orbital 
radius, $\theta$ is the pulsar's ecliptic latitude and $d$ its distance 
(\cite{rt91a}).  Thus, a timing precision of $\sim$ 1$\mu$s limits detection 
of $\pi$ to pulsars \lapp 1 kpc away.

The upper panel of Figure 1 shows the timing residuals of PSR 
J1744$-$1134 folded with a semi-annual period for a model fit
excluding parallax. Also plotted is the timing signal corresponding to
$\pi=$ 2.8 mas --- the best fit value for parallax. The residuals for a
model fit including $\pi$ are shown in the lower panel. The former
fit resulted in a post-fit residual rms of 0.80$\mu$s, while the
latter produced an rms of 0.47$\mu$s.  Thus, the inclusion of parallax
in the timing model yields a significant improvement to its accuracy.
There are some indications of red noise in the TOAs in Figure 1.
These systematics arise from inaccuracies in the calibration of pulse
profiles, and introduce red noise in the TOAs on time-scales of hours
(\cite{bri99}).  Such short-term systematics average out over the
semi-annual time-scale of parallax, and should not affect our measured
value.  Although the statistical error in $\pi$ from the fit is 0.1
mas, the presence of systematics in the residuals will tend to cause 
an underestimation of this error. To quantify the effects of systematics 
on our parallax error we divided our data into 4 sub-sets of 1 year. 
Fits to independent pairwise combinations of these sub-sets were consistent 
with $\pi=$ 2.8$\pm$0.3 mas.  To this precision timing noise does not 
contribute to the measured parallax. The 8 parameter values resulting from 
the best fit, along with derived parameters, are presented in Table 2.

\section{Discussion}

Our parallax measurement of PSR J1744$-$1134 places it at a distance
of  357$^{+43}_{-35}$ pc. This distance is over twice the value of 166 pc 
derived from the DM using the Taylor \& Cordes (1993) model. The improved  
distance measurement implies a mean electron density in the path 
to the pulsar of $n_{e}=$ (8.8$\pm$0.9)$\times$10$^{-3}$ cm$^{-3}$. This is  
the lowest $n_{e}$ for any known pulsar with a low $z$-height. To determine whether 
this LOS is exceptional, we have compared $n_{e}$ for PSR J1744$-$1134 with 
those of another 11 local ($d$\lapp 1 kpc) pulsars with independent distance 
estimates. With the exception of PSRs J1024$-$0719, B1534+12, 
and B0833$-$45 (Vela) all such estimates were derived from parallax measurements 
(Table 1). A firm upper distance limit to PSR  J1024$-$0719 of 226 pc results 
from assuming that its observed  spin-down  rate is entirely due to its transverse 
motion (Shklovskii effect;  Shklovskii 1970)\nocite{shk70}. Note that this
upper limit is $\sim$40\% lower than the DM  derived distance. A
distance to  PSR B1534+12 of 1.1$\pm$0.2 kpc (consistent  with their
parallax  measurement) has been derived by Stairs \etal (1998)
\nocite{sac+98}  from a measured change in its orbital period derivative  
under the assumption that general relativity is correct. Recently, Cha, Sembach 
\& Danks (1999)\nocite{csd99} derived a distance to the Vela supernova remnant 
of 250$\pm$30 pc --- half the canonically accepted value (\cite{mil68}). 

PSRs J1744$-$1134 and J1024$-$0719 are two of only three isolated MSPs
detected at X-ray energies.  Since these MSPs have similar spin
parameters (\cite{tsb+99}), and presumably similar evolutionary
histories, comparing their X-ray properties is useful in understanding
the origin of their X-ray emission. With our revised distance estimates,
the X-ray luminosity of PSR J1024$-$0719,  $L_{x} <$ 1$\times$10$^{29}
d^{2}$erg s$^{-1}$, is less than a third of  the previously accepted
value, and for PSR J1744$-$1134, $L_{x}=$ 4$\times$10$^{29} d^{2}$erg 
s$^{-1}$, is slightly higher than the previous value, and much 
larger than that of PSR J1024$-$0719 (cf. \cite{bt99}).

When the distances to the sample of 12 local pulsars are projected onto
the Galactic plane a striking asymmetry in the electron distribution
becomes evident. Figure 2 depicts their projected positions and 
LOS electron densities. It is apparent that the LOS electron densities to 
pulsars in the third Galactic quadrant are systematically higher than those 
in the first. In general, densities in quadrants 1 and 3 are respectively lower 
and higher than values expected  from the Taylor \& Cordes model. Most of the 12 
pulsars are close to and above the Galactic plane ($z$-heights $<$200 pc, $b>$0). 
PSRs J1713+0747 and B1534+12 are exceptional ($z=$ 473 and 822 pc  respectively) 
and therefore are expected to have low electron densities. PSRs J0437$-$4715, 
B0833$-$45 and PSR B1451$-$68 are the only pulsars at negative latitudes. 
The asymmetry may be more pronounced since  the upper parallax limit of PSR B1929+10 
sets an upper limit to its electron density ($n_{e}<$ 0.013 cm$^{-3}$), while 
the  Shklovskii effect implies a firm lower bound to the $n_{e}$ of PSR J1024$-$0719 
($n_{e}>$ 0.029 cm$^{-3}$).

The Taylor \& Cordes model for the Galactic free electron distribution
is generally reliable for distances \gapp 1 kpc. In the local region,
the model electron density is uniform with $n_{e}=$ 0.02 cm $^{-3}$
in the Galactic plane; it takes no account of the peculiar properties of
the LISM. Independent distance limits to local pulsars are crucial 
to constraining the electron density in this region. With such constraints 
it may be possible to re-evaluate the DM derived distances to nearby pulsars. 
To further characterize the distribution of electrons in the LISM it 
is useful to relate their location to other interstellar features, such as 
bubbles, superbubbles, and clouds of neutral gas.

There is strong evidence for an elongated cavity in the neutral component 
of the LISM. This cavity surrounds the Sun and extends several hundred 
parsecs into quadrant 3 (\cite{luc78}). The cavity appears as a region 
of low reddening extending  500 pc between $l=$ 210\arcdeg$\,$ and 
255\arcdeg$\,$ and 1.5 kpc toward $l=$  240\arcdeg$\,$. Running counter 
to this is very heavy obscuration  beyond $\sim$100 pc in the first
quadrant.  Similarly, H{\small I} column densities derived from ultraviolet  
observations show a marked paucity in H{\small I} along LOSs directed towards 
$l=$ 230\arcdeg$\,$(\cite{fy83}; \cite{par84}). A similar morphology for this 
cavity is gleaned from Na{\small I} absorption measurements (\cite{wcv+94}). 
Figure 2 shows the position of the 12  pulsars with respect to the cavity in 
reddening material and H{\small I}. We note that pulsars with the highest electron 
column densities are located close to or within this cavity.

There are several features of interest within this cavity. One of these is 
the local hot bubble (LHB): a volume encompassing the Sun distinguished by low 
neutral gas densities and a 10$^{6}$ K, soft X-ray emitting gas (for 
reviews see \cite{cr87}; \cite{bre96}). The size of the LHB may be 
estimated by assuming that its extent in any given direction is proportional 
to the intensity of soft X-rays from that part of the sky (\cite{cr87};
\cite{scc90}). A revision of this `displacement model' based on {\it
ROSAT} observations indicates that the LHB only partially fills the
cavity (\cite{sef+98}). In some regions (e.g., quadrant 3) the boundary is 
not well defined because of enhancements in the X-ray emission from material 
thought to be not local. In this model the electron  density inside the bubble 
is $n_{e}\sim$ 0.005 cm$^{-3}$.  Recently Heiles  (1998b) \nocite{hei98b} has 
provided evidence from H{\small I}, infrared,  radio continuum and 0.25 keV 
X-ray observations for a superbubble in quadrant 3.  This superbubble is 
centered at a distance of $\sim$0.8 kpc in the direction of $l=$ 
238\arcdeg$\,$ --- its nearest and farthest boundaries  are about 0.2 
and 1.3 kpc distant, respectively. Figure 2 shows that the local hot
bubble and the superbubble fill much of the cavity, although the extent of 
the region between them is not well defined.

Electrons within the LHB contribute only a small amount to the DMs of local 
pulsars. Therefore, the large number of pulsars with lower than expected electron 
densities in quadrant 1 implies a dearth of ionized material in this quadrant at 
least out to $\sim$ 1 kpc. This has implications for models of the 
Galactic magnetic field. The majority of rotation measurements (RMs) have been made 
to pulsars in quadrant 1. An analysis of RMs by Rand \& Lyne (1994)\nocite{rl94} 
has the uniform component of the local magnetic field  directed towards 
$l \sim$ 90\arcdeg$\,$ with a magnitude of $\sim$1.4 $\mu$G. About 400 pc towards 
$l=$ 0\arcdeg$\,$ the field direction reverses. Calibration of nearby pulsar 
distances is indispensable to mapping the large-scale RM changes that distinguish 
this effect. 

The very high electron column densities toward PSRs B0823+26 
(0.055 cm$^{-3}$), B0833$-$45 (0.270 cm$^{-3}$), B0950+08 (0.023 cm$^{-3}$) 
and J1024$-$0719 ($n_{e}>$ 0.029 cm$^{-3}$) in quadrant 3 indicate the 
presence of dense ionized gas immediately beyond the LHB. Alternatively, the 
LHB may have a much higher $n_{e}$, as in the model of Breitschwerdt
\& Schmutzler (1994).\nocite{bs94} This could account for the nearby
excess of electrons in quadrant 3 but would require a further
reduction in electron density in quadrant 1. If the dense ionized  
material is outside the LHB, the relative deficiency of electrons along  
the LOS to PSR J0437$-$4715 may be accounted for if the gas is clumped or 
has a non-uniform $z$-distribution. There is independent evidence for the 
existence of ionized clouds in this region. Gry, York \& Vidal-Madjar (1985)
\nocite{gyv85} and more recently Dupin \& Gry (1998) \nocite{dg98}
have  investigated the properties of highly ionized clouds along LOSs to 
$\beta$ Canis Majoris ($l$, $b$, $d=$ 226\arcdeg$\,$, $-$14\arcdeg$\,$, 153 pc). 
Gry \etal (1985) concluded  that H{\small II} with $n_{e}=$ 0.07-0.14 cm$^{-3}$ 
fills 40-90 pc of the $\beta$ CMa LOS. Dupin \& Gry (1998) find that two clouds 
dominate this LOS. They have also postulated that these clouds are in the process 
of cooling and recombining after having been shocked and ionized by some violent 
event. It is possible that this highly ionized region is associated with an 
interaction between the LHB and the superbubble.

\section{Conclusion}

We have measured a parallax of 2.8$\pm$0.3 mas for PSR J1744$-$1134
based  on analysis of pulse arrival time data. The corresponding
distance of  357$^{+43}_{-35}\,$pc is more than twice that derived from 
the DM using the Taylor \& Cordes (1993) electron density model. 
The derived electron density, $n_{e}=$ (8.8$\pm$0.9)$\times$10
$^{-3}$ cm$^{-3}$, is the  lowest of any pulsar close to the Galactic
equator. We compared the LOS electron density to PSR J1744$-$1134 with 
those to another 11 nearby pulsars with independent distance estimates. 
We conclude that there is a striking asymmetry in the distribution of 
electrons in the LISM, with electron densities for pulsars in the 
third Galactic quadrant systematically higher than those in the first. 
We speculate that these ionized regions are associated with interactions 
at the boundary between the LHB and the more distant superbubble.

The authors thank the referee, whose insightful comments contributed to 
the clarity of this paper.

\newpage
\section*{Figure Captions}

\noindent

Figure 1.  Timing residuals for PSR J1744$-$1134. The upper panel shows residuals 
for frequencies centered at 0.66 (triangles) and about 1.4 GHz (circles), after fitting 
a timing model excluding parallax. The residuals have been folded at a period of 0.5 years 
and the sinusoid corresponds to the form and magnitude of a parallax of 2.8 mas.    
The lower panel displays the residuals after fitting for this parallax. 

\noindent
Figure 2. Features in the local ISM projected on to the Galactic plane. The Sun is 
located at the origin with concentric annuli at 200 pc intervals. The position 
of 12 pulsars are marked by circles, with radii proportional to the mean electron 
density in the path to the pulsar. The exceptionally large $n_{e}$ for Vela 
is marked with a circled `V' centered on its position. Letters refer to the pulsars listed 
below the figure, and electron densities (in units of cm$^{-3}$) are quoted alongside each 
pulsar. The positive and negative signs superposed on C and E points indicate lower and 
upper bounds, respectively. Note the asymmetric distribution of electron density. Contours 
plotted as solid lines indicate color excess (Lucke 1978; Figure 5). The innermost contour 
level is $E(B-V)=$ 0.1 and the levels increase outward in steps of 0.1 mag. 
Depressed contours are indicated by tick marks. The contour plotted as a dashed line 
indicates the neutral hydrogen column density with a level of $N$(H{\small I})$=$ 5$\times$10$^{19}$ 
cm$^{-2}$ (Frisch \& York 1983; Figure 1c). The cross-section of the LHB through 
the Galactic plane is shown as a shaded patch at the Sun's location (Snowden \etal 1998; Figure 11). 
The superbubble identified by Heiles (1998b) is plotted as a shaded ellipse. We note that the 
high electron densities correlate with the position of the H{\small I} cavity and with the region 
between the LHB and the superbubble.

\newpage
\begin{deluxetable}{lccccc}
\tablecolumns{3}
\tablewidth{0pc}
\tablecaption{Parallax and Distance to 12 Pulsars}
\tablehead{
\colhead{Pulsar}& \colhead{$\pi$}& \colhead{l}& \colhead{b}& \colhead{Distance}& \colhead{Reference}\\ 
\colhead{}& \colhead{mas}& \colhead{}& \colhead{}& \colhead{pc}& \colhead{} \\}
\startdata
J0437--4715 & 5.6$\pm$0.8   & 253.4 & $-$41.96 & 179$^{+29}_{-23}$ &      8     	\\
B0823+26    & 2.8$\pm$0.6   & 197.0 &    31.7  & 357$^{+97}_{-63}$ &      6	        \\
B0919+06    & 0.31$\pm$0.14 & 225.4 &    36.4  & 3200$^{+2700}_{-1000}$&  5     	\\
B0950+08    & 7.9$\pm$0.8   & 228.9 &    43.7  & 127$^{+14}_{-12}$ &      6      	\\
B1451--68   & 2.2$\pm$0.3   & 313.9 &  $-$8.5  & 455$^{+70}_{-56}$ &      2     	\\
B1534+12    & $<$1.7        &  19.9 &    48.3  & $>$588            &      9     	\\
J1713+0747  & 0.9$\pm$0.3   &  28.8 &    25.2  & 1111$^{+556}_{-278}$ &   4 		\\
J1744--1134 & 2.8$\pm$0.3   &  14.8 &     9.2  & 357$^{+43}_{-35}$ & \nodata    	\\
B1855+09    & 1.1$\pm$0.3   &  42.3 &     3.1  & 909$^{+341}_{-195}$&     7     	\\
B1929+10    & $<$4 \tablenotemark{a} & 47.4 & $-$3.9     & $>$250  &      1      	\\
B1937+21    & $<$0.28       &  57.5 &  $-$0.3  & $>$3571 & 	          7      	\\
B2021+51    & 0.95$\pm$0.37 &  87.9 &     8.4  & 1050$^{+250}_{-150}$&    3     	\\
\enddata
\tablenotetext{a}{A measurement of 21.5$\pm$0.3 mas by  Salter, Lyne, \& Anderson (1979)
\nocite{sla79} was not included in our analysis.\nocite{sla79}}
\tablerefs{(1) \cite{bs82}; (2) \cite{bmk+90a}; 
(3) \cite{cbs+96}; (4) \cite{cfw94}; (5) \cite{fgb+99}; (6) \cite{gtwr86}; 
(7) \cite{ktr94}; (8) \cite{sbm+97}; (9) \cite{sac+98}}
\end{deluxetable}

\newpage
\begin{deluxetable}{ll}
\tablecolumns{2}
\tablewidth{0pc}
\tablecaption{Observed and Derived Parameters for PSR J1744--1134}
\tablehead{
\colhead{Parameter}&
\colhead{Value \tablenotemark{a}  }   \\ }
\startdata                                                    
R. A. (J2000) & 17$^{\rm h}$ 44$^{\rm m}$ 29\fs390963(5) \\
Decl.  (J2000) & $-11\arcdeg$ 34\arcmin~54\farcs5746(5) \\
Proper motion in R.A. (mas y$^{-1}$) & 18.72(6) \\
Proper motion in Decl. (mas y$^{-1}$) & $-$9.5(4)  \\
Annual parallax (mas) & 2.8(3) \\
Period, $P$  (ms)      & 4.07454587512695(3) \\
Period derivative, $\dot P$ ($10^{-20}$) & 0.89405(9) \\
Epoch of period and position (MJD)  & 50434.0 \\
Dispersion Measure (cm$^{-3}$ pc) & 3.1388(3)\\
Timing data span (MJD) & 49730 -- 51209 \\
RMS timing residual ($\mu$s) & 0.47 \\
Number of timing points  &  158 \\
Galactic latitude \& longitude (l, b) &  14.79, 9.18  \\
Parallax distance  (pc)   & 357$^{+43}_{-35}$\\
$DM$ derived distance (pc) & 166  \\ 
$\dot P$ upper distance limit  (pc) & 1910 \\
Composite proper motion (mas y$^{-1}$) & 21.0(2)\\
Celestial position angle of proper motion (deg) & 116.8(9) \\
Transverse Velocity (km s$^{-1}$) & 36(4) \\
\enddata
\tablenotetext{a}{Figures in parentheses represent 1$\sigma$ uncertainties in the last
digit quoted}
\end{deluxetable}

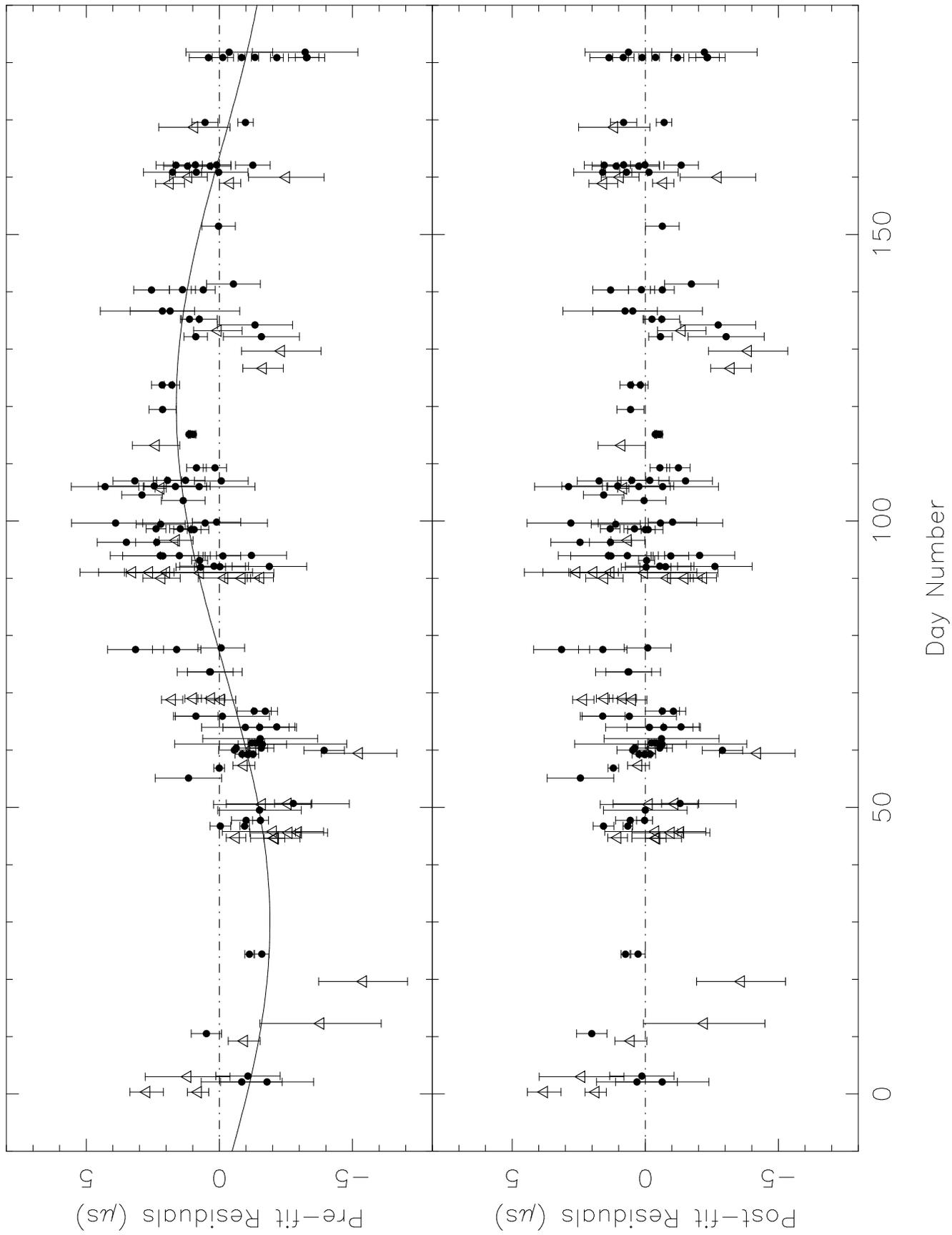
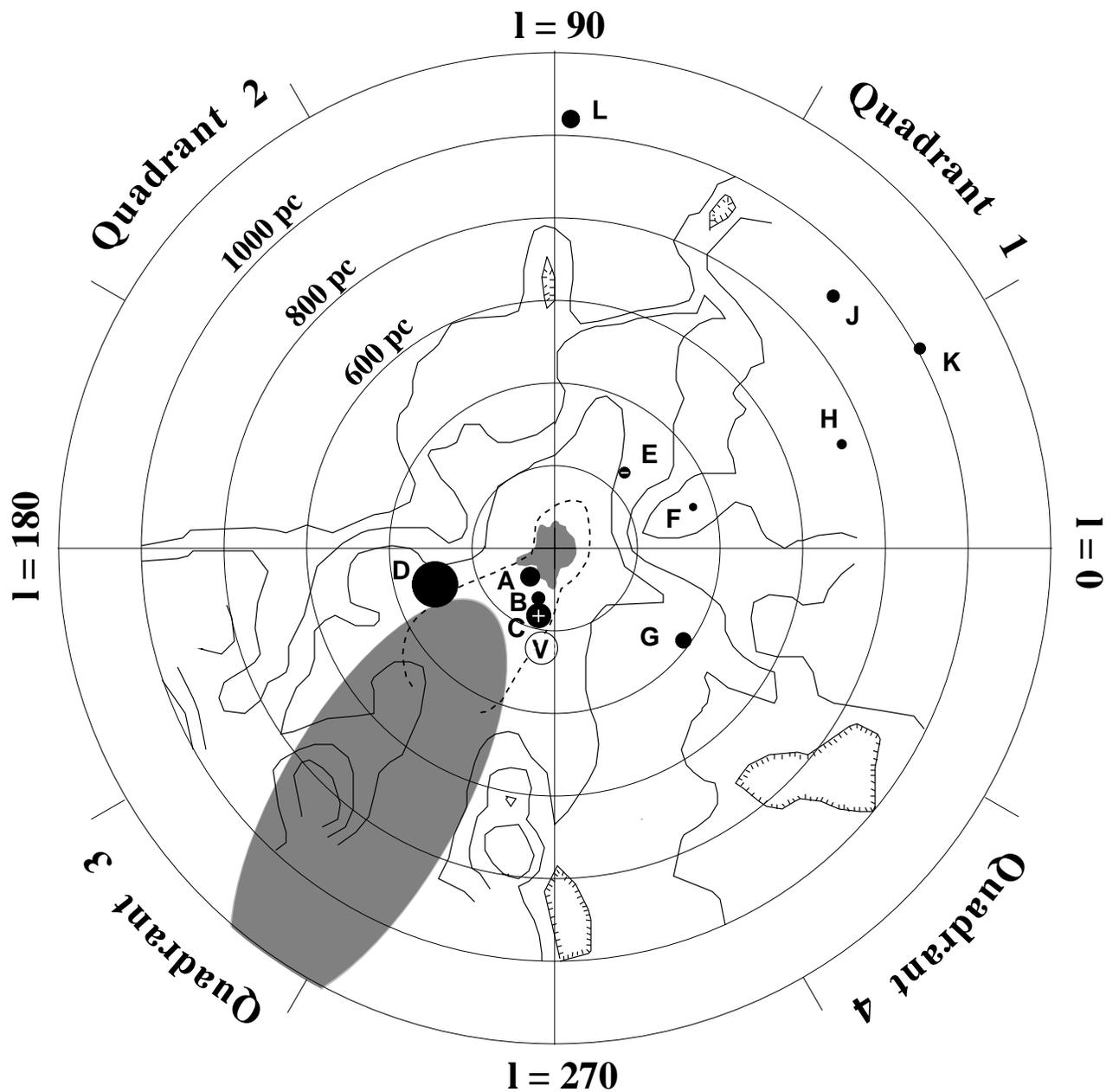
\end{document}